\documentclass[a4paper,11pt]{article}
\begin{document}
\parindent=0pt
\def\NPB#1#2#3{Nucl. Phys. {\bf B} {\bf#1} (#2) #3}
\def\PLB#1#2#3{Phys. Lett. {\bf B} {\bf#1} (#2) #3}
\def\PRD#1#2#3{Phys. Rev. {\bf D} {\bf#1} (#2) #3}
\def\PRL#1#2#3{Phys. Rev. Lett. {\bf#1} (#2) #3}
\def\PRT#1#2#3{Phys. Rep. {\bf#1} C (#2) #3}
\def\ARAA#1#2#3{Ann. Rev. Astron. Astrophys. {\bf#1} (#2) #3}
\def\ARNP#1#2#3{Ann. Rev. Nucl. Part. Sci. {\bf#1} (#2) #3}
\def\MODA#1#2#3{Mod. Phys. Lett. {\bf A} {\bf#1} (#2) #3}
\def\NC#1#2#3{Nuovo Cim. {\bf#1} (#2) #3}
\def\ANPH#1#2#3{Ann. Phys. {\bf#1} (#2) #3}
\def\PTP#1#2#3{Prog. Th. Phys. {\bf#1} (#2) #3}
\def\JHEP#1#2#3{JHEP. {\bf#1} (#2) #3}
\begin{titlepage}
\thispagestyle{empty} 
\begin{large}
\title{Can One Phase Induce All CP Violations\\ Including Leptogenesis?}
\end{large}
\author{{\bf Yoav
  Achiman}\footnote{e-mail:achiman@post.tau.ac.il}\\
  [1.5cm]
  School of Physics and Astronomy, Tel Aviv University, 69978 Tel Aviv,
  Israel\\
  Department of Physics, University of Wuppertal, D--42097 Wuppertal,
  Germany}

\date{ March 2004}
 
\maketitle
\setlength{\unitlength}{1cm}
\begin{picture}(5,1)(-12.5,-10)
\put(0,0.5){TAUP 2768-2004}
\end{picture}
\begin{abstract}
In the framework of a SUSY $SO(10)$ model a phase is generated spontaneously
 for the
$B-L$ breaking VEV. Fitting this phase to the observed CP violating K,B
decays all other CP breaking effects are uniquely predicted. In particular, 
the amount of Leptogenesis can be explicitly calculated and found to be in 
the right range and sign for the BAU.
\vspace{90pt}

{\it PACA numbers: 11.30.Fs, 12.10.Dm, 12.15.Hh, 13.35.Dx1}  
\end{abstract}
\thispagestyle{empty}
 
\end{titlepage}
\parindent=0pt

CP violation is directly observed only in the decays of the K and B mesons. The
present experimental results~\cite{exp_cp} are consistent at the moment
with the standard model (SM). I.e. CP breaking is induced by a phase in the 
Cabibbo, Kobayashi, Maskawa (CKM) mixing matrix of the quarks.\\
Extensions of the SM using right-handed (RH) neutrinos, that account for
the neutrino oscillations, involve in general phases which allow for CP 
violation
in the leptonic sector also. This CP breaking is difficult to observe but
may be detected as soon as neutrino factories are available. 
The observation of neutrino-less double beta decays may be also an indication
for Majorana phases in the neutrino sector~\cite{beta}.

Spontaneous generation of baryon asymmetry in the universe (BAU) needs
CP violation~\cite{Sakharov}. It is clear now that it requires also 
extension of the SM, while baryon asymmetry in the universe (BAU) a la 
Fukugita and Yanagida~\cite{FY} due to leptogenesis~\cite{lepgen} is 
the most popular and promising theory for the BAU.\\

Where is the CP breaking coming from? \\
CP breaking can be induced via phases in the Yukawa coupling,
in the interactions of the LH and RH gauge bosons and in the VEVs.
Phases in the spontaneously generated VEVs lead naturally to 
violation of CP. This spontaneous breaking can also help to solve the 
strong CP problem~\cite{nelson}~\cite{barr}.\\
The spontaneous violation of CP was already suggested long ago by 
T.~D.~Lee~\cite{Lee}. In the framework of $SO(10)$ GUT
spontaneous breaking was first discussed by Harvey, Reiss and Ramond~
\cite{HRR}.
Recently, Bento and Branco~\cite{bento} added to the SM a heavy Higgs scalar 
with a $B-L$ violating VEV to generate spontaneous CP violation.\\  

In general, the known  CP violation in the hadronic sector is not
related to the leptonic one. Even the CP breaking needed for leptogenesis
is usually independent of that in the leptonic sector. Hence, CP 
violation in the leptonic sector is in general not predictable. 
Predictability can be gained only in terms of a specific model.
There are quite a few models relating CP violation in the neutrino sector
to leptogenesis~\cite{lep-lep} but no conventional SUSY GUT which connects 
the leptogenesis to the observed violation in the K and B decays is 
presently known.\\

I would like to suggest in this paper that the one and only origin for CP 
violation is a spontaneous breaking at high energies. A phase in the 
$B-L$ breaking VEV
can induce all manifestations of CP violation. This phase can be fixed by
the observed breaking in the K and B decays and the other CP
violations are then predicted. In particular, we will show explicitly that
 within a SUSY $SO(10)$ model the amount and sign of leptogenesis are as is
needed to have the right BAU.\\

Let me first show how a phase can be spontaneously generated in the $SU(5)$
singlet component of a scalar ${\bf 126}$ representation of $SO(10)$.
It was already pointed out by Harvey, Ramond and Reiss ~\cite{HRR} that
there is a natural way to 
break CP spontaneously at high energies. This is due to the
fact that $({\bf 126})^4$ is $SO(10)$ invariant.
${\bf \Phi}_{\overline{\bf 126}}$ is the Higgs representation used to break 
down $B-L$ . Its $SU(5)$ singlet component gives also masses to the heavy
RH neutrinos. The corresponding large VEV induces also small VEVs in the
components of ${\bf\Phi}_{\overline{\bf 126}}$ that transform like ${\bf 2}_L$
under the SM~\cite{Magg} which play a role in the light fermion mass 
matrices.\\

Assume that all the parameters in the $SO(10)$ invariant Lagrangian are real.
If the three fermionic families are in ${\bf \Psi}_{\bf 16}$'s,  only 
${\bf \Phi}_{\bf 10}$, ${\bf \Phi}_{\overline{\bf 126}}$ and
${\bf \Phi}_{\bf 120}$ can contribute to the mass terms:

\begin{equation}
{\bf 16}\times {\bf 16}=({\bf 10}\oplus{\bf 126})_S\oplus({\bf 120})_{AS}
\end{equation}

Suppose we have chosen global symmetries that dictate a (super-)potential 
of the form\footnote{Note that ${\bf 10}$ is a real 
representation.}~\cite{bento}

\begin{equation}
V(\lambda_1,\lambda_2, ....)=V_0+[...+\lambda_1({\bf \Phi}_{\bf 10})^2_S][
({\bf \Phi}_{\bf 126})^2_S + ({\bf \Phi}_{\overline {\bf 126}})^2_S]+ 
\lambda_2 [({\bf \Phi}_{\bf 126})^4_S + ({\bf \Phi}_{\overline {\bf 126}})^4_S] \end{equation}

and that those are the only phase dependent terms after the spontaneous
breaking\footnote{
For a detailed discussion of possible scalar potentials see Ref.~\cite{HRR}.
The $[({\bf \Phi}_{\bf 126})^4_S + ({\bf \Phi}_{\overline{\bf 126}})^4_S]$ 
part serves
also to break the continuous global symmetries avoiding massless  Nambu-
Goldston bosons.}. 
If the $SU(5)$ singlet component of ${\bf \Phi}_{\bf 126}$,
${\bf\Phi}_{\overline{\bf 126}}$
acquire a VEV as well as the right component of ${{\bf \Phi}_{\bf 10}}$:
\begin{equation}
<{\bf \Phi}_{\bf 10}>=\frac{\upsilon}{\sqrt 2}
\qquad\qquad <{\bf \Phi}_{\bf 126}>
=\frac{\Upsilon}{\sqrt 2} e^{i\alpha}\qquad\qquad
\end{equation}

The phase dependent part of the potential can be then written as
\begin{equation}
V(\upsilon,\Upsilon,\alpha)=A\cos(2\alpha)+B\cos(4\alpha).
\end{equation}

For B positive and $|A|>4B$ the absolute minimum of the potential is 
obtained with
\begin{equation}
\alpha=\frac{1}{2}\arccos(\frac{A}{4B}).
\end{equation}

This spontaneous generation of a phase in the large VEV $\Upsilon$, will
generate also phases in the induced small VEVs which give mass to the light
fermions. Those will lead to CP violation in the quark and lepton sectors.
The value of the spontaneously generated phase $\alpha$ depends on arbitrary
parameters in the Higgs potential. Its actual value can be however fixed
by the requirement that the phases of the induced light VEVs will give the
observed CP violation in the K,B decays. All other manifestations of CP 
violation will be then explicitly predicted.  In particular the amount and
sign
of leptogenesis are    then predicted in models where $M^{\scriptscriptstyle 
Dirac}_\nu$ is known.\\

Let me now explicitly calculate the amount of leptogenesis in a SUSY $SO(10)$
model where a phase is generated spontaneously in the $B-L$ breaking VEV.
The model was developed in a series of papers~\cite{model}~\cite{richter}.
 It was
originally aimed to find explicitly the mixing angles which are hidden 
in the SM, like RH rotations. Those allow to calculate explicitly e.g.  
the proton decay branching ratios as well as all mass matrices and in 
particular the Dirac neutrino mass matrix and the RH neutrino mass matrix
which are needed for the calculation of the leptogenesis. We will use here
the mass matrices given in Ref.~\cite{richter}. This is a renormalizable SUSY
$SO(10)$ model i.e. $B-L$ is broken via 
${\bf \Phi}_{\overline{\bf 126}}+{\bf \Phi}_{{\bf 126}}$ while 
${\bf \Phi}_{\overline{\bf 126}}$ gives mass to the RH neutrinos 
(without using non-
renormalizable contributions). The origin of CP breaking in
the model is a phase in the $SU(5)$ singlet component of one ${\bf \Phi}_
{\overline{\bf  126}}$. A global horizontal symmetry $U(1)_F$
dictates the asymmetric Fritzsch texture~\cite{NNI} for the fermionic mass 
matrices
and the possible VEVs in the different Higgs representations. By fitting the
free parameters to the observed masses and CKM matrix a set of non-linear 
equations is obtained. These equations have five solutions which obey all the
restrictions, i.e. five sets of explicit mass matrices. The Dirac neutrino
mass matrices have the texture:

\begin{equation}
M^{\scriptscriptstyle Dirac}_{\nu}= \left(
\begin{array}{ccc}
0&A&0\\
B&0&C\\
0&D&E
\end{array} \right).
\end{equation}

They are given explicitly in Appendix I.\\
The RH neutrino mass matrices have the following form in our model: 

\begin{equation}
M_{\nu_R}=e^{i\alpha} \left(
\begin{array}{ccc}
0&a&0\\
a&0&0\\
0&0&-b
\end{array} \right) M_{R} .
\end{equation}
Where the real $a,b>0$.
The corresponding eigenmasses are given in Table~1.\\


Table 1. The masses of the RH neutrinos for the five solutions in 
$10^{13}$~GeV.

\begin{center}
\begin{tabular}{|c|c|c|c|c|c|}
\hline
Solution $10^{13}$GeV&1&2&3&4&5\\
& &&&& \\[-10pt]
\hline           
& &&&& \\[-10pt]
$M_1=M_2$&5.2&9.1&16&18&12\\
& &&&& \\[-10pt]
\hline
& &&&& \\[-10pt]
$M_3$&$8\times5.2$&$7\times9.1$&$3\times16$&$3\times18$&$2\times12$\\
& &&&& \\[-3pt]
\hline
\end{tabular}
\end{center}
\vspace{15pt}
What is Leptogenesis?\\
Out of equilibrium CP violating decays of RH neutrinos, $N_i$,
produce excess of the lepton number $\delta L\neq 0$ . This will induce baryon 
asymmetry through $B+L$ conserving sphaleron processes~\cite{FY}~\cite
{lepgen}~\cite{Buch}.\\
The amount of CP violation in these decays is:
$$
\epsilon_i=\frac{\Gamma(N_i \rightarrow L_i + \Phi)\quad - \quad
\Gamma({N_i}^\dagger \rightarrow {L_i}^\dagger + {\Phi}^\dagger)}
{\Gamma(N_i \rightarrow L_i + \Phi)\quad + \quad \Gamma({N_i}^\dagger
\rightarrow {L_i}^\dagger + {\Phi}^\dagger)}
$$ 
Knowing the details of CP violation in the leptonic sector as well as the 
RH mixing angles\footnote{Note, that $M^{\dagger}M$ is diagonalized using the
RH mixing matrix.}, one is able to calculate explicitly    the BAU via 
leptogenesis. This is the main test of the model.\\

Let us denote the Dirac neutrino mass matrix $M^{\scriptscriptstyle Dirac}_\nu$ in the basis
where $M_{\nu_R}$ is real diagonal with positive eigenvalues: $M_{D}$.
In this basis $\epsilon_i$ can be expressed as follows
$$
\epsilon_i={\frac{1}{8\pi v^2 (M_D^{\dagger}M^{}_D)_{11}}}{\sum_{j\neq 1}}Im
[{(M_D^{\dagger}M^{}_D)^2_{ij}}]f(M_j^2/M_i^2)
$$
where
$$
f(x)=\sqrt{x} [ln(1+\frac{1}{x})+\frac{2}{x-1}] 
$$
and $v=174\times\sin\beta$ GeV~\footnote{$\tan\beta=10$ is used in the model
~\cite{richter}}.\\

$M_{\nu_R}$ is given in eq.(7) and its eigenmasses in Table 1.\\

It is diagonalized by a matrix $U$
$$
U^TM_{\nu_R}U={diag}(M_1,M_2,M_3)=M_3\ \ {diag}(\frac{M_1}{M_3},
\frac{M_1}{M_3},1)
$$
$$
U=OP\qquad {{\hbox {where}}}\qquad P=e^{-({i/2})\alpha} {diag}(i,1,i)
$$
and
$$
O= \left(
\begin{array}{ccc}
\frac{1}{\sqrt{2}}&\frac{1}{\sqrt{2}}&0\\
-{\frac{1}{\sqrt{2}}}&\frac{1}{\sqrt{2}}&0\\
0&0&1
\end{array} \right).
$$

\vspace{1cm}
In this basis, in terms of eq. (6)\\

$$
M^{\dagger}_DM^{}_D=\left(
\begin{array}{ccc}
{1/2}(|A|^2+|B|^2+|D|^2)&{i/2}(|A|^2-|B|^2+|D|^2)&{1/\sqrt{2}(B^\dagger C-
D^\dagger E})\\
-{i/2}(|A|^2-|B|^2+|D|^2)&{1/2}(|A|^2+|B|^2+|D|^2)&{i/\sqrt{2}(B^\dagger C+
D^\dagger E})\\
{1/\sqrt{2}(B C^\dagger-D E^\dagger })&{-i/\sqrt{2}(B C^\dagger+
D E^\dagger})&|C|^2+|E|^2
\end{array} \right).
$$
\vspace{20pt}

This gives the following general results
\begin{equation}
Im((M^{\dagger}_DM^{}_D)_{12} (M^{\dagger}_DM^{}_D)_{12})=
Im((M^{\dagger}_DM^{}_D)_{21} (M^{\dagger}_DM^{}_D)_{21})=0
\end{equation}
$$
(M^{\dagger}_DM^{}_D)_{11}=(M^{\dagger}_DM^{}_D)_{22}
$$
Due to the degeneracy of $N_1,N_2$,  the decay of both contributes to 
$\epsilon_i$. However, eq. (8) avoids the possible singularity in $f(x)$. 
Hence,
$$
\epsilon_L={\frac{1}{8\pi v^2 (M_D^{\dagger}M^{}_D)_{11}}}(Im
[(M_D^{\dagger}M^{}_D)^2_{13}]+Im[(M_D^{\dagger}M^{}_D)^2_{23}]) f(M_3^2/M_1^2).
$$ 
The BAU is given then (in the minimal supersymmetric SM) as
$$
Y_B=-{1/3}\frac{\epsilon_L}{g^*} d_{B-L}
$$
where $g^*=228.75$ and $d_{B-L}$ is the dilution factor due to inverse decay 
washout effects and lepton number violating scattering. It must be obtained 
by solving the corresponding Boltzmann equation. There are different 
approximate solutions in the literature. The frequently used approximate 
solution~\cite{Kolb} is good only for 
$$
K=\frac{{\tilde m_1} M_P}{1.7 \times 8\pi v^2 \sqrt{g^*}}=
\frac{{\tilde m_1}\quad(eV)}{1.08\times 10^{-3}\quad( eV)}>1
$$
where $\qquad{\tilde m_1}=\frac{(M^{\dagger}_DM^{}_D)_{}11}{M_1}\quad$.
In our model however, $\qquad K \approx 10^{-2}\quad$ .\\
Buchm\"uller, Di Bari and Pl\"umacher~\cite{Buch} studied recently 
in detail both
cases $K>1$ and $K<1$. They found that for $K<1$ one must take into account 
thermal corrections ue to the gauge bosons and the top quark. Hence, $d_{B-L}$
depends on ``initial conditions'' and they found that for $K\approx 10^{-2}$~
\footnote{See Figure 9 in their paper where $d_{B-L}$ is called 
$\kappa_f$.}
$$
10^{-4}\geq d_{B-L}\leq 10^{-2}
$$

Hirsch and King~\cite{king} give  empirical approximate solutions for
the case \\ $K\ll1$~. The solution corresponding to our model is
$$
Log_{10}(d_{B-L})=0.8\times Log_{10}({\tilde m_1~eV}) + 1.7 +
 0.05\times Log_{10}(M_1/10^{10}~GeV).$$ 
I will use this expression to have a definite prediction. The results for
the five solutions are given in Table 2.\\

Table 2. The CP asymmetry $\epsilon_L$, the dilution factor $d_{B-L}$
and the Baryon asymmetry $Y_{B}$ for the five solutions.
\begin{center}
\begin{tabular}{|c|c|c|c|}
\hline
Solution&$ \epsilon_L$& $d_{B-L}$&$Y_B$ \\
\hline 
& & &\\ [-10pt]
1& $-6.5\times10^{-7}$&0.0064&$6.1\times10^{-12}$ \\
& && \\ [-10pt]
\hline 
& && \\ [-10pt]
2& $-6.6\times10^{-5}$& 0.0074&$7.1\times10^{-10}$ \\
& && \\ [-10pt]
\hline
& && \\ [-10pt]
3& $-7.4\times10^{-5}$&0.0088&$9.5\times10^{-10}$ \\
& && \\ [-10pt]
\hline
& && \\ [-10pt]
4& $-1.3\times10^{-6}$&0.009&$1.7\times10^{-11}$\\ 
& && \\[-10pt]
\hline
& && \\[-10pt]
5&$-5.6\times10^{-5}$&0.06&$4.9\times 10^{-10}$\\
& && \\[-1pt]
\hline
\end{tabular}
\end{center}

This must be compared with the experimental results:\\
BOOMerANG and DASI~\cite{BOOM}
$$
0.4\times 10^{-10}\leq Y_{B} \geq 1.0\times 10^{-10}
$$
WMAP and Sloan Digital Sky Survey~\cite{WMAP}
$$
Y_{B}=(6.3\pm 0.3)\times 10^{-10}
$$
Hence:\\
{$\bullet$} Solutions 1 and 3 are probably excluded. The other solutions are 
consistent with the experimental observation, especially if the uncertainty 
in $d_{B-L}$ is taken into account.\\
{$\bullet$}  All solution have the right sign. This is the main prediction of
the model in view of the uncertainty in $ d_{B-L}$.\\
I must emphasize that there is no ambiguity in the prediction of the sign
because of the following reasons:\\  
a) The sign of $M_1$ must be positive because $\epsilon_i$ is calculated in 
terms of $M_D$ which is the neutrino Dirac mass matrix in the basis where 
the RH neutrino mass matrix (7) is diagonal, real and positive.\\
b) The parameters and especially the phases of $M^{\scriptscriptstyle Dirac}_
\nu$ (6) are fixed without ambiguity for each one of the above solutions, 
although one cannot write explicitly their dependence on $\alpha$. As was
mentioned before, the entries to the mass matrices are solutions of non-
linear equations in which the induced components of ${\bf \Phi}_{\overline
{\bf 126}}$ (with the phase $\alpha$) are involved. The physical value of
$\alpha $ is then fixed by requiring that $J_{Jarlskog}\sim 10^{-5}$ to be
$\alpha \sim 0.003$\footnote{In a recent paper Frampton, Glashow and Yanagida
in ref.~\cite{lep-lep} presented a model where the sign of the BAU can be 
related to the CP violation in neutrino oscillation experiments. In our model
both CP violation in the neutrino oscillation as well as the sign of the BAU
are predicted in terms of CP violation in the quark sector.}.  \\

To complete the predictions of the model let me use the
complex lepton mixing matrix $U_{PMNS}$ of Ref.~\cite{richter}
(see Appendix II) to give
the amount of CP violation in the neutrino oscillation
$$
J_{leptons}= Im(U_{11}U_{22}U^*_{12}U^*_{21})
$$\quad 
and the value of $<m_{ee}>$ 
$$
<m_{ee}>=\sum_{i=1}^{3} (U_{e1})^2 m_i
$$
relevant for the neutrino-less double-beta decay $\beta\beta_{0\nu}$
\footnote{${m_1}$ in our solutions is of $O(10^{-3}eV)$.}.
 See Table 3.\\

Table 3. The CP violation invariant for the leptonic sector $J_{leptons}$
and the effective neutrino mass for the neutrino-less double-beta decay 
for the five solutions.
$$
\begin{tabular}{|c|c|c|c|c|c|}
\hline
Solution&1& 2&3&4&5\\
& &&&& \\[-10pt]
\hline
& &&&& \\[-10pt]
$J_{leptons}$&0.0092&0.000059&9.8$\times 10^{-6}$&7.8$\times 10^{-6}$&6.6$\times 10^{-6}$\\
& &&&& \\
\hline
& &&&& \\
$<m_{ee}>$&0.0031&0.005&0.0068&0.0056&0.0029\\
\hline
\end{tabular}
$$

\vspace{10pt}
CONCLUSIONS\\

I presented in the paper the following observations:\\
CP is naturally broken spontaneously at high energies in $SO(10)$ GUTs.\\
A phase is generated in a VEV and not in the Yukawa couplings, as it is 
usually done.
This can be used as the only origin CP violation.\\
In the framework of a SUSY $SO(10)$ model that uses this idea, fitting
to the observed CP violation, as it is reflected in the CKM matrix,
fixes uniquely the CP breaking in the leptonic sector without free parameters.
An explicit calculation of leptogenesis in this model gives solutions 
consistent with the range and sign of the observed BAU\footnote{``A 
common origin for all CP violations'' was suggested recently also by
Branco, Parada and Rebelo~\cite{Branco}.   They use a non-SUSY SM extended
by adding scalar Higgs, leptons and exotic vector-like quarks. The complex 
phase is generated spontaneously in the VEV of the heavy singlet scalar
meson. The connection with the low energy CP violation in the hadronic 
sector is obtained only via mixing with the exotic quarks. They give also 
no explicit value for the leptogenesis.}.\\
Our model applies the conventional see-saw mechanism~\cite{SS}, it is possible
however, to use a similar program for the type II see-saw~\cite{II} as 
well~\cite{YA}.\\
The large value of the RH neutrino mass can be incompatible with the 
gravitino problem if SUSY is broken in the framework of mSUGRA.
Possible solutions are discussed in the literature. E.g. Ibe, Kitano, 
Murayama   and Yanagida~\cite{grav} presented very recently a nice 
solution based on anomaly mediated SUSY breaking.\\

\vspace{200pt}

APPENDIX I\\

\vspace{-10pt}
The Dirac neutrino mass matrices 
for the five solutions 
(for $\tan \beta= 10$) in GeV.\\ 


\begin{center}
\begin{tabular}{|c|c|c|c|c|c|}
\hline
Solution GeV&1&2&3&4&5\\
& &&&& \\[-10pt]
\hline
& &&&& \\[-10pt]
Re$(M^{\scriptscriptstyle Dirac}_\nu)_{12}$&17.486 &26.953 &-41.320 &
-41.320 &-28.274\\
& &&&& \\[-10pt]
\hline
& &&&& \\[-10pt]
Im$({M}^{\scriptscriptstyle Dirac}_\nu)_{12}$&0.0394&0.0607&0.0929&-0.0929&-0.06356\\
& &&&& \\[-10pt]
\hline
& &&&& \\[-10pt]
Re$(M^{\scriptscriptstyle Dirac}_\nu)_{21}$&17.654&27.120&-41.218&-41.218&-28.172\\
& &&&& \\[-10pt]
\hline
& &&&& \\[-10pt]
Im$({M}^{\scriptscriptstyle Dirac}_\nu)_{21}$&0.0394&0.0607&0.0929&-0.0929&-0.06356\\
& &&&& \\[-10pt]
\hline
& &&&& \\[-10pt]
$(M^{\scriptscriptstyle Dirac}_\nu)_{23}$&-113.425&-142.425&116.073&82.073&102.073\\
& &&&& \\[-10pt]
\hline
& &&&& \\[-10pt]
$(M^{\scriptscriptstyle Dirac}_\nu)_{32}$&-14.700&14.302&10.695&44.695&24.695\\
& &&&& \\[-10pt]
\hline
& &&&& \\[-10pt]
Re$(M^{\scriptscriptstyle Dirac}_\nu)_{33}$&-127.913&-176.670&146.103&146.103&78.715\\
& &&&& \\[-10pt]
\hline
& &&&& \\[-10pt]
Im$({M}^{\scriptscriptstyle Dirac}_\nu)_{33}$&-0.3152&-0.4249&0.2788&0.2788&0.1271\\
& &&&& \\[-1pt]
\hline
\end{tabular}
\end{center}
\vspace{350pt} 

APENDIX II\\

\vspace{5mm}
The leptonic mixing matrix for the different solutions.
\vspace{2cm}
\begin{center}
\begin{tabular}{|c|c|c|c|c|c|}
\hline
Solution &1&2&3&4&5\\
& &&&& \\[-10pt]
\hline
& &&&& \\[-10pt]
Re$(U_{PMNS})_{11}$&-0.8583&0.8136&0.7465&0.8579&0.8740\\
& &&&& \\[-10pt]
\hline
& &&&& \\[-10pt]
Im$(U_{PMNS})_{11}$&0.000004&0.00034&-0.000001&-0.000001&0.000001\\
& &&&& \\[-10pt]
\hline
& &&&& \\[-10pt]
Re$(U_{PMNS})_{12}$&-0.5104&-0.5778&-0.6589&-0.5059&-0.4806\\
& &&&& \\[-10pt]
\hline
& &&&& \\[-10pt]
Im$(U_{PMNS})_{12}$&-0.000007&0.000007&-0.00027&-0.00021&-0.0002\\
& &&&& \\[-10pt]
\hline
& &&&& \\[-10pt]
Re$(U_{PMNS})_{13}$&-0.0526&-0.0644&0.0927&0.0897&0.0717\\
& &&&& \\[-10pt]
\hline
& &&&& \\[-10pt]
Im$(U_{PMNS})_{13}$&0.000002&0.00026&0.00042&0.00004&0.00003\\
& &&&& \\[-10pt]
\hline
& &&&& \\[-10pt]
Re$(U_{PMNS})_{21}$&-0.3496&-0.4869&-0.4653&-0.3754&-0.2492\\
& &&&& \\[-10pt]
\hline
& &&&& \\[-10pt]
Im$(U_{PMNS})_{21}$&0.00191&0.00190&0.00212&0.0017&0.00088\\
& &&&& \\[-10pt]
\hline
& &&&& \\[-10pt]
Re$(U_{PMNS})_{22}$&0.6567&-0.6168&-0.6167&-0.7364&-0.5670\\
& &&&& \\[-10pt]
\hline
& &&&& \\[-10pt]
Im$(U_{PMNS})_{22}$&-0.0030&0.0029&0.00260&0.0031&0.00018\\
& &&&& \\[-10pt]
\hline
& &&&& \\[-10pt]
Re$(U_{PMNS})_{23}$&-0.6682&-0.6185&-0.6350&-0.5628&-0.7829\\
& &&&& \\[-10pt]
\hline
& &&&& \\[-10pt]
Im$(U_{PMNS})_{23}$&0.0031&0.00285&0.0029&0.0026&0.0028\\
& &&&& \\[-10pt]
\hline
& &&&& \\[-10pt]
Re$(U_{PMNS})_{31}$&-0.3756&-0.3176&-0.4756&-0.3508&-0.4172\\
& &&&& \\[-10pt]
\hline
& &&&& \\[-10pt]
Im$(U_{PMNS})_{31}$&0.00082&0.00085&0.00216&0.0009&0.0011\\
& &&&& \\[-10pt]
\hline
& &&&& \\[-10pt]
Re$(U_{PMNS})_{32}$&0.5552&-0.6168&-0.4309&-0.4492&-0.6664\\
& &&&& \\[-10pt]
\hline
& &&&& \\[-10pt]
Im$(U_{PMNS})_{32}$&-0.00121&0.00127&0.0009&0.00097&0.0014\\
& &&&& \\[-10pt]
\hline
& &&&& \\[-10pt]
Re$(U_{PMNS})_{33}$&0.7421&0.7832&0.7669&0.82168&0.6179\\
& &&&& \\[-10pt]
\hline
& &&&& \\[-10pt]
Im$(U_{PMNS})_{33}$&-0.00163&-0.00204&-0.0020&-0.00213&-0.0016\\
\hline
\end{tabular}
\end{center}

\end{document}